\documentclass{article}
\usepackage{amsmath} 
\usepackage{epsfig} 
\usepackage{cite} 

\input{epsf} 
\newcommand\as{\alpha_{{S}}}

\def\b0{b_0}

\def\beq{\begin{equation}} 
\def\eeq{\end{equation}} 
\def\beeq{\begin{eqnarray}} 
\def\eeeq{\end{eqnarray}} 
 
\def\to{\rightarrow}
 
\def\nn{\nonumber}

\def\ms{${\overline {\rm MS}}$}

\def\gev2{{\rm GeV}^2}
\def\nn{\nonumber}

\def\lapproxeq{{\ \lower 0.6ex \hbox{$\buildrel<\over\sim$}\ }}
\def\gapproxeq{{\ \lower 0.6ex \hbox{$\buildrel>\over\sim$}\ }}
\def\tb{\mbox{tan $\beta \,$}}

\def \be  {\begin{equation}}
\def \ee  {\end{equation}}
\def \ba  {\begin{eqnarray}}
\def \ea  {\end{eqnarray}}
\def \baa {\begin{eqnarray*}}
\def \eaa {\end{eqnarray*}}
\def \bb  {\begin{thebibliography}}
\def \eb  {\end{thebibliography}}

\begin{document} 

\begin{titlepage}
\renewcommand{\thefootnote}{\fnsymbol{footnote}}
\begin{flushright}
     \end{flushright}
\par \vspace{10mm}

\begin{center}
{\Large \bf
Soft-gluon resummation for pseudoscalar Higgs boson production 
at hadron colliders.}
\end{center}
\par \vspace{2mm}
\begin{center}
{\bf Daniel de Florian\footnote{email address: deflo@df.uba.ar} and 
Jos\'e Zurita\footnote{email address: jzurita@df.uba.ar}}\\

\vspace{5mm}

Departamento de F\'\i sica, FCEYN, Universidad de Buenos Aires,\\
(1428) Pabell\'on 1 Ciudad Universitaria, Capital Federal, Argentina\\
\vspace{2mm}

\vspace{5mm}

\end{center}

\par \vspace{2mm}
\begin{center}

 {\large \bf Abstract}\\
\vspace{1cm}

We compute the threshold-resummed cross section for pseudo-scalar 
MSSM Higgs boson production by gluon fusion at hadron colliders. 
The calculation is performed at next-to-next-to leading logarithmic accuracy. 
We present results for both the LHC and  Tevatron Run II. We analyze the factorization and renormalization scale dependence of the results, finding that after performing the resummation the corresponding cross section can be computed with an accuracy better than 10\%.

 \end{center}
\begin{quote}
\pretolerance 10000
                    
\end{quote}

\vspace*{\fill}
\begin{flushleft}
October 2007

\end{flushleft}
\end{titlepage}

\setcounter{footnote}{1}
\renewcommand{\thefootnote}{\fnsymbol{footnote}}

\section{Introduction}
\label{sec:intro}

The hunt for the Standard Model (SM) Higgs boson is clearly one of the biggest physics 
goals at the LHC. 
Its search is not only a great challenge from the experimental point of view, it also requires a huge theoretical effort in order to provide very precise 
predictions both for signal and background. Besides the SM, there are many other options for New Physics (NP).
Among 
the possible scenarios, the MSSM is a 
 promising one. It provides a solution for the hierarchy problem and also introduces a good dark matter candidate, the lightest supersymmetric particle. 
The Higgs boson sector of this model consists in two complex Higgs doublets. 
After the EWSB, three degrees of freedom
are absorbed by the electroweak massive bosons and 
the remaining five give rise to the usual SM scalar Higgs boson ($h$),
a heavier neutral one ($H$), two charged scalars ($H^{\pm}$) 
and a pseudoscalar neutral Higgs ($A$). 
In the MSSM, the tree level masses depend upon two parameters, which can be 
selected to be the mass of the pseudoscalar Higgs ($m_A$)
and the ratio of the vacuum expectation values of the two doublets 
\tb $= v_2/ v_1$ \cite{Spira:1993bb,Djouadi:1992pu}.
These parameters are constrained by LEP and Tevatron experiments, with bounds given by
$m_A > 100 $ GeV  and with the interval $0.5 < \,$tan $ \beta < 2.4$ currently excluded \cite{Schael:2006cr}.
The actual status of SM and SUSY Higgs production at LHC is thoroughly reviewed in Ref.~\cite{Harlander:2006pc}.

The main channel for the production of an on-shell neutral Higgs boson at the LHC, both for SM and MSSM bosons, is the gluon fusion process. Therefore, the corresponding cross section needs to be under good theoretical control. The calculation, 
for both CP-even and CP-odd Higgs particles, has been 
performed up to next-to-leading order (NLO) in an exact way \cite{Spira:1995rr}. Surprisingly, the NLO corrections turned out to be quite large, accounting for an increase of almost 100 \% of the leading-order (LO) result. Furthermore, adding the NLO term did not result in a sizeable reduction of the renormalization and factorization scale dependence. Nevertheless, the NLO corrections showed that the gluon-Higgs interactions could be very well approximated by using an effective Lagrangian approach, inspired by low energy theorems. This effective theory is obtained by considering an infinite mass limit for the top quark \cite{heavymt}. This approximation allowed to perform the next-to-next-to leading order (NNLO) calculations for the inclusive cross section and even to analyze the transverse momentum distribution of the Higgs boson \cite{tm,tmres}.

At NNLO accuracy, the dominant soft-virtual corrections were first computed in \cite{Catani:2001ic,Harlander:2001is} , while the full result \cite{Ravindran:2003um,Anastasiou:2002yz,Harlander:2002wh,Anastasiou:2005qj} became available later.
For pseudoscalar Higgs, the full NNLO contribution was computed in Refs. \cite{Ravindran:2003um,Anastasiou:2002wq,Harlander:2002vv}.
These corrections turned out to be considerably smaller than the NLO ones, bringing a reasonable reduction in the scale dependence as well.

The soft-gluon resummation, which includes in the cross section the dominant effects of the higher order terms of the perturbative QCD series, was performed only for the scalar Higgs, 
up to next-to-next-to- leading logarithmic (NNLL) accuracy \cite{Catani:2003zt}. These corrections showed a modest numerical impact, and a further (slight) reduction of the scale dependence, providing a hint on the convergence of the asymptotic QCD perturbative series. The quantitative reliability of the soft-gluon approximation was probed by comparing the truncation of the resummed result at NLO and NNLO with the fixed-order calculation \cite{Catani:2001ic,Catani:2003zt}. This comparison showed that the resummed result can also be trusted away from the threshold region, thus convalidating the accuracy of the threshold resummation in the entire kinematical phase space.

 In the present work we compute the cross section for CP-odd 
Higgs production by gluon fusion up to NNLL, 
in the infinite top mass limit. As the corrections for scalar and pseudoscalar Higgs do not greatly differ, we still expect, based on the scalar Higgs result, that the soft-gluon resummation is a highly accurate approximation to the exact result in the whole phase space, and not only around the threshold region.

This article is organized as follows. In Section 2 we introduce our notation and briefly review the 
theoretical framework, including QCD cross sections at fixed-order and the basics of soft-gluon resummation. In Section 3 we present and discuss our results, both for the LHC and the Tevatron Run II while in Section 4 we give our conclusions.

\section{Theoretical frame}
\label{sec:review}
According to the mass factorization theorem, the inclusive cross section for 
the production of the pseudoscalar 
Higgs boson by the collision of hadrons $h_1$ and $h_2$ may be written as
\begin{align}
\label{had}
\sigma(s,M_A^2) =& 
\sum_{a,b} \int_0^1 dx_1 \;dx_2 \; f_{a/h_1}(x_1,\mu_F^2) 
\;f_{b/h_2}(x_2,\mu_F^2) \int_0^1 dz \;\delta\!\left(z -
\frac{\tau_A}{x_1x_2}\right) \nn \\
& \cdot {\hat \sigma}_{ab}(\hat{s},M_{A}^2) (z;\as(\mu_R^2), 
M_A^2/\mu_R^2;M_A^2/\mu_F^2) \;,
\end{align}
where $M_A$ is the pseudoscalar Higgs boson mass, $\tau_A=M_A^2/s$,
and $\mu_F$ and $\mu_R$ are the factorization and 
renormalization scales, respectively. ${\hat \sigma}_{ab}$ denote 
the partonic cross section for the process $a+b \to A+X$, computable 
in perturbative QCD.
The parton densities of the colliding hadrons are denoted by 
$f_{a/h}(x,\mu_F^2)$, where the subscript $a$ labels the parton type. 
We use parton densities as defined in the \ms\ factorization scheme.
For practical purposes, one works with the hard coefficient functions 
${G}_{ab}$ instead of the partonic cross sections, the first one given by
\beq
{\hat \sigma}_{ab}=\sigma^{(0)} \, z \, {G}_{ab}(z) \, \, ,
\eeq
where $\sigma^{(0)}$ is the Born level contribution.
The incoming massless partons $a,b$ do not couple to the pseudoscalar Higgs boson 
directly. In hadron collisions the main production mechanism is through 
heavy quark triangle loops and therefore, the total cross section also depends 
on the top ($M_t$) and bottom ($M_b$) quark masses.

The NLO coefficients $G_{ab}^{(1)}$ have been exactly computed in 
Ref.~\cite{Spira:1995rr}, where it was also observed that the NLO Higgs boson 
cross section can be well approximated in the low \tb regime by considering its limit $M_t \gg M_A$ 
\cite{Dawson:1990zj}.
Hence, along this  paper we will work within the large-$M_t$
approximation: we consider the case of a single heavy quark (the top),
and $N_f=5$ light-quark flavors, neglecting all the contributions to 
$G_{ab}^{(n)}$ that vanish when $M_A/M_t \to 0$. Nevertheless, the full 
dependence on $M_t$ and $M_b$ is included in $\sigma^{(0)}$ in order 
to improve the accuracy of the calculation.
The large-$M_t$ approximation allows the use of the effective-Lagrangian 
approach \cite{heavymt,Chetyrkin
,Kramer:1998iq}, that shrinks the top quark triangle loop into an effective 
point-like vertex, considerably simplifying the evaluation of the 
Feynman diagrams involved in the process.

The heavy top mass limit is reliable as long as tan$^{2} \, \beta\ll M_t / M_b$
\cite{Field:2002pb}, because of the quark-Higgs couplings. Therefore, any 
calculation relying in the $M_t \to \infty$ approximation is valid for low values of \tb. Moreover, we assume       
that the lighter squarks are much heavier than the top quark, thus their contribution can be safely neglected.

The coefficient function $ G_{ab}(z)$ is dominated by soft terms in the limit $z \to 1$. Therefore, our main objective is to study the effect of soft-gluon contributions to 
all perturbative orders. This task requires to work in the Mellin 
(or $N$-moment) space \cite{Sterman:1986aj,CatTrenres}. We thus
introduce our notation in the $N$-space.

We consider the Mellin transform $\sigma_N(M_A^2)$ of the hadronic cross
section $\sigma(s,M_A^2)$. The $N$-moments with respect to $\tau_A=M_A^2/s$
at fixed $M_A$ are customarily  defined as follows:
\begin{equation}
\label{sigman}
\sigma_N(M_A^2) \equiv \int_0^1 \;d\tau_A \;\tau_A^{N-1} \;\sigma(s,M_A^2) 
\;\;.
\end{equation} 
In $N$-moment space, Eq.~(\ref{had}) takes a simple factorized form
\begin{equation}
\label{hadn}
\sigma_{N}(M_A^2) =\sum_{a,b}
\; f_{a/h_1, \, N}(\mu_F^2) \; f_{b/h_2, \, N}(\mu_F^2) 
\; {\hat \sigma}_{ab,\, N}(\as(\mu_R^2), M_A^2/\mu_R^2;M_A^2/\mu_F^2) \;,
\end{equation}
where $f_{i/h, \, N}$ and ${\hat \sigma}_{ij,\, N}$ represent the  Mellin 
transforms of the parton distributions $f_{i/h}$ and of the partonic 
cross sections  ${\hat \sigma}_{ij}$, respectively. 

In order to perform the threshold resummation, we first note that the 
threshold region $z \to 1$ corresponds to the limit $N \to \infty$ in Mellin 
space. The dominant contribution in this limit is due to the large logarithmic terms $\as^n \ln^mN$ .
Being the only channel open at the Born level, the $gg\rightarrow A$ subprocess is the unique partonic contribution that can give rise to the large threshold logarithms. The formalism to systematically perform soft-gluon resummation for hadronic 
processes was developed in Refs.~ \cite{Sterman:1986aj,CatTrenres}. In the case of Higgs boson production, one has
\begin{equation}
\label{ggscaling}
G_{gg, \,N} = \as^2 \left\{ 1 + \sum_{n=1}^{+\infty} \as^n 
\sum_{m=0}^{2n} G_A^{(n,m)} \ln^mN \right\}+ {\cal O}(1/N)
= {G}_{gg,\, N}^{{\rm (res)}} + {\cal O}(1/N) \;\;,
\end{equation}
where the dominant contributions in the large-$N$
limit may be reorganized in the following {\em all-order} resummation formula:
\begin{align} 
\label{resformula} 
{G}_{gg,\, N}^{{\rm (res)}}(\as(\mu_R^2), M_A^2/\mu_R^2;M_A^2/\mu_F^2) 
&=\as^2(\mu_R^2)\, C_{gg}(\as(\mu^2_R),M_A^2/\mu^2_R;M_A^2/\mu_F^2) \nn \\ 
&\cdot \exp \{ {\cal G}_h(\as(\mu^2_R), \ln N;M_A^2/\mu^2_R,M_A^2/\mu_F^2)\}
\; . 
\end{align}
The function $C_{gg}(\as)$ contains all the contributions that are constant 
in the large-$N$ limit. They are originated from the hard virtual contributions 
and non-logarithmic soft corrections, and can be computed as a power series 
expansion in $\as$. The large logarithmic terms  $\as^n \ln^mN$ 
(with $1 \leq m \leq 2n$), which are due to soft-gluon radiation,
are included in the exponential factor $\exp{\cal G}_h$. It can be expanded as 
\begin{align} 
\label{gexpan} 
~\vspace{-.5cm} {\cal G}_h\!\left(\as(\mu^2_R),\ln N;\frac{M_A^2}{\mu^2_R}, 
\frac{M_H^2}{\mu_F^2}\right)
&= \ln N \; g_h^{(1)}(\lambda) + 
g_h^{(2)}(\lambda, M_A^2/\mu^2_R;M_A^2/\mu_F^2 )  \nonumber \\ 
&+ \as(\mu^2_R) 
\;g_h^{(3)}(\lambda,M_A^2/\mu^2_R;M_A^2/\mu_F^2 ) 
+ {\cal O}(\as(\as \ln N)^k)  
\end{align} 
where $\lambda=\b0 \as(\mu^2_R)\ln N$ and $\b0$ is the first coefficient 
of the QCD $\beta$-function.

Due to the universality of the soft-gluon emission, the ${\cal G}$ factor is independent on the type of Higgs boson produced in the final state. Hence, the coefficients of the expansion are the same for both $h$ and $A$. The expressions for these coefficients can be found in Ref. \cite{Catani:2003zt}. The difference in the resummed expansion between the scalar and pseudoscalar Higgs only shows up in the $C_{gg}$ factor which, being partially originated on the hard-virtual contribution, obviously depends on the Higgs type under consideration. For the sake of brevity, we only write the differences between the $A$ and the $h$ terms, which read
\beq
\begin{split}
 \Delta C_{gg}^{(1)}=C_{gg,A}^{(1)}-C_{gg,h}^{(1)}&=\frac{1}{2} \\
 \Delta C_{gg}^{(2)}=C_{gg,A}^{(2)}-C_{gg,h}^{(2)}&= 
\frac{1939}{144}+3 \gamma_E^2+\pi^2-\frac{21}{16} N_f - (\frac{19}{8} + \frac{N_f}{3}) \, \ln \frac{M_A^2}{M_t^2}+(\frac{3}{8} - 3 \gamma_E -\frac{5}{6} N_f) \, \ln \frac{M_A^2}{\mu_F^2} \\  &+(-\frac{33}{8}-\frac{1}{4} N_f) \, \ln \frac{M_A^2}{\mu_R^2} \, \, ,
\end{split}
\eeq
where $N_f$ is the number of different light quark flavors.

Going back to Eq.~(\ref{gexpan}), 
the term $\ln N \; g_h^{(1)}$ resums all the {\em leading} logarithmic (LL) 
contributions
$\as^n \ln^{n+1}N$, $g_h^{(2)}$ contains the {\em next-to-leading} logarithmic 
(NLL) terms $\as^n \ln^{n}N$, $\as g_h^{(3)}$ collects
the {\em next-to-next-to-leading} logarithmic (NNLL) terms 
$\as^{n+1} \ln^{n}N$, and so forth.
In this context, the product $\as \ln N$
is formally considered as being of order unity. Therefore, the ratio of two
successive terms in the expansion (\ref{gexpan}) is formally of 
${\cal O}(\as)$, which makes the resummed logarithmic expansion in 
Eq.~(\ref{gexpan}) as systematic as the usual fixed-order expansion 
in powers of $\as$. 

The leading collinear contributions can also be included in the soft-gluon resummation formula by performing \cite{Catani:2003zt,Kramer:1998iq} the following shift
\beq
C_{gg}^{(1)} \to  C_{gg}^{(1)} + 6 \frac{\ln N}{N}
\eeq
that correctly resums all the terms of the type  $(\as^n \ln^{2n-1}N)/N$ that appear in $G_{gg}^{(n)}$ .

When attempting for the resummation, one is interested in taking some advantage of the full fixed-order cross section calculation as well. It is therefore customary \cite{Catani:2003zt} to perform a matching between both approaches, which can be schematically written as
\beq
\sigma^{matched}=\sigma^{res}+\sigma^{f.o}-\sigma^{res} \arrowvert_{f.o}
\eeq
where $\sigma^{res}$ corresponds to the result obtained using Eq.~(\ref{resformula}), $\sigma^{f.o}$ is the fixed-order cross section and $\sigma^{res} \arrowvert_{f.o}$ represents the expansion of the resummed result at the same order in $\as$ as the fixed-order result. 
This improved \emph{matched} cross section is our final result for the process. In consequence, throughout the paper we shall refer to the different orders of the matched cross sections directly as LL, NLL and NNLL. The accuracy of the matching is assured by the order at which the $C_{gg}$ coefficient is computed, being obtained by a direct comparison between the resummed and the fixed-order calculation. Therefore, in $\sigma^{f.o}-\sigma^{res} \arrowvert_{f.o}$ only the hard terms, which are strongly suppressed in the $N \to \infty$ limit, survive.

Recently, the function $g_h^{(4)}$ was presented \cite{Moch:2005ba,Moch:2005ky}. In principle, it allows to perform the resummation up to NNNLL accuracy. However, the full matched calculation at NNNLL, can only be done if the fixed-order NNNLO result \footnote{At least, the full soft-virtual contributions are necessary to compute the coefficient $C_{gg}^{(3)}$.} were available. Nowadays, only the soft contribution was derived \cite{Moch:2005ky}. 
We have decided to look at the effect of including the
$g_h^{(4)}$ term and setting $C^{(3)}_{gg}=0$. This inclusion leads to a really slight modification of the full NNLL result 
thus validating the 
convergence of the resummed series expansion, and allowing us to safely neglect the $g_h^{(4)}$ function along this work.

Finally, once the expression of the cross section has been computed in $N$-space, the physical result can be obtained by Mellin inversion. In order to avoid the Landau singularities explicitly present in the exponential factor in Eq.~(\ref{gexpan}) we use the \emph{Minimal Prescription}, as described in Ref. \cite{Catani:1996yz}.

\section{Results}

We have developed the program THIGRES, a FORTRAN code to compute
the resummed (fixed-order) cross section up to NNLL (NNLO) accuracy, both for 
scalar and pseudoscalar Higgs boson in the heavy $M_t$ limit.
The improvement over previous calculations lies in the fact that the 
partonic cross sections up to NNLO are directly written in Mellin space. 
The Mellin coefficients of the hard functions $G_{ab}^{(n)}$ were presented 
in \cite{Blumlein:2005im}. We have recomputed these coefficients in an 
analytical way, by performing the Mellin transform of the results presented in Refs \cite{Ravindran:2003um,Anastasiou:2002yz,Harlander:2002wh,Anastasiou:2002wq,Harlander:2002vv}, finding some non-negligible differences in the $gg$ and  $q \bar{q}$ channels at NNLO with respect to \cite{Blumlein:2005im}. 
The essential ingredients for Mellin transformation can be found in \cite{Mellingredients}.
 We have taken advantage of the ANCONT Fortran code
\cite{Blumlein:2000hw} which provides most of the required special functions 
in $N$-space.

One essential missing element to tackle the calculation in Mellin space are the PDFs. The available parton distributions are always given in the $x$ space. In order to transform them to $N$-space, we first perform a fit of the densities at the needed scale using a functional form that allows for a simple analytical Mellin transform. We find that a linear combination of 
Eulerian functions $x^{(\alpha)} (1-x)^{\beta}$, which in Mellin space give rise to beta functions $B(\alpha+N,\beta-1)$ are enough to reproduce all the features of the usual parton distributions  \cite{deFlorian:2005yj}. After performing some clever sampling, the result of the fit allows to compute $N$-moments analytically with an accuracy better than 0.5\%. Having both the necessary fixed-order and resummed coefficients and the PDFs in Mellin space, one is able to perform the calculation in the most efficient way with a considerable reduction in the required computer time and a gain in precision.

We have worked with the MRST set of PDFs; 
using the 2001 LO \cite{Martin:2001es} and the 2002 NLO 
and NNLO \cite{Martin:2002dr}, although the code THIGRES allows the use of 
other PDFs. For the presentation of our results we use 
$M_t=176$ GeV and $M_b=4.75$ GeV. Therefore, the cross section for the pseudoscalar 
Higgs boson is reliable only if \tb $<6$\footnote{We have explicitly checked the accuracy of the 
infinite top mass approximation by comparing our NLO results with the exact 
NLO 
ones provided by the FORTRAN code HIGLU \cite{Spira:1995mt}. 
Within this bounds the accuracy is always better than 10 percent. }.

\begin{figure}[!htp]
      \begin{center}
        \epsfig{figure=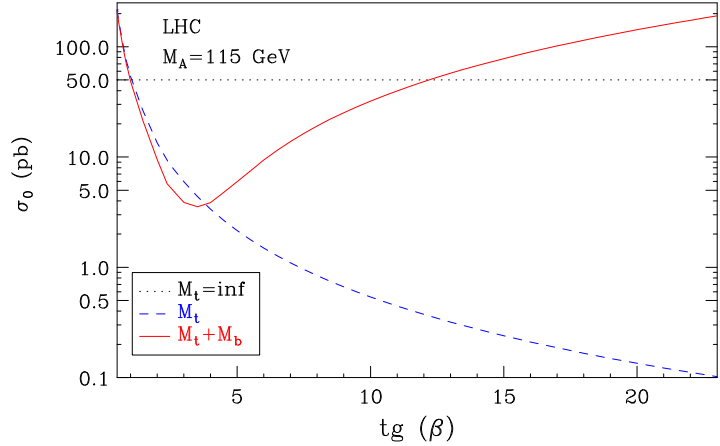,angle=0,height=6cm}
	\end{center}
        \caption{\label{fig:borntgb}{\em \tb dependence of the Higgs 
production Born cross section at the LHC for $M_A=115$ GeV, in the 
infinite top mass limit (dots), including top mass effects 
(dashes) and both top and bottom mass effects (solid) }.}
     \end{figure}

In order to check the validity range the approximation in \tb, we will start studying the dependence of the results upon this parameter. As we are neglecting the finite top mass effects in the partonic cross sections, \tb only appears in the Born term $\sigma_0$. 
In Fig.~\ref{fig:borntgb} we plot $\sigma_0$ as a function of \tb, showing the corresponding variation of the Born cross section, for a $M_A=115$ GeV boson at the LHC, according to whether one includes top and/or
bottom mass effects. 
It is clear that the use of the infinite top mass limit in the Born cross section is not reliable. At least the $M_t$ dependence must be included in $\sigma_0$. 
The solid (top + bottom) and dashed (only top) curves are similar around \tb $=1$; for this particular value, the pseudoscalar Higgs coupling to up and down type quarks is the same as for the scalar one. The inclusion of the bottom mass can be neglected until one enters the region where \tb becomes close to $\sqrt{M_t/M_b}$\footnote{At this particular value, the $A$-top coupling is equal to the $A$-bottom coupling.}. Near this point, both curves start to separate from each other. It is rather noticeable that for values of \tb $\ge 10$
the bottom effects are completely dominant, and therefore any calculation relying on the infinite top mass approximation can not be trusted for those values. This behaviour is certainly  expected, since the couplings of the pseudoscalar Higgs boson to the up (down) type quarks are suppressed (enhanced) by a factor \tb.

\begin{figure}[h]
      \begin{center}
        \epsfig{figure=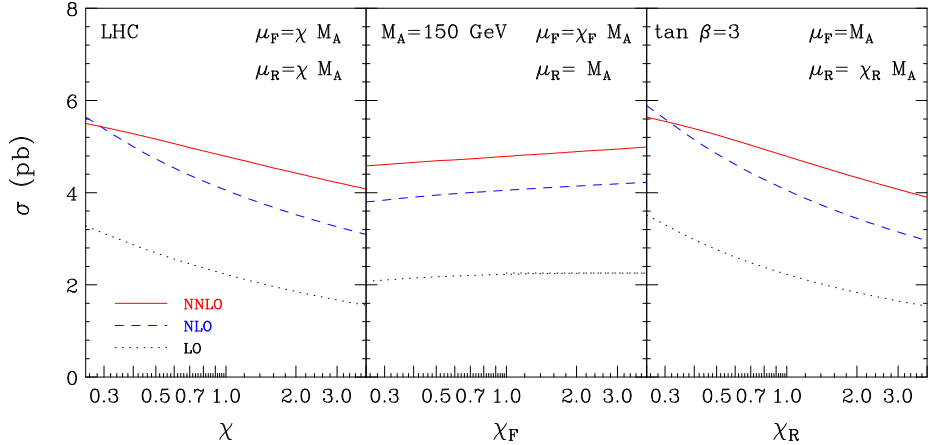,angle=0,height=6cm}
	\end{center}
        \caption{\label{fig:lhc150}{\em Scale dependence of the Higgs 
production  cross section at the LHC for $M_A=150$ GeV, \tb $=3$, at 
LO (dots), NLO (dashes) and NNLO (solid). }}
     \end{figure}

It is very important to know the dependence of the cross section upon the renormalization and factorization scales, as a way to estimate the size of the higher order terms not yet included in the perturbative expansions and therefore evaluate the uncertainties on the theoretical calculations.
This dependence is shown in 
Fig.~\ref{fig:lhc150}, considering a $M_A=150$ GeV Higgs boson at the LHC, using \tb $=3$, for the fixed-order LO, NLO and NNLO results. 
In this figure, both scales were varied from $M_A/4$ to $4 \, M_A$ in
three different ways.
In the plot on the left, the varying scales were chosen to be equal 
($\mu_F=\mu_R=\chi M_A$). 
In the plot on the center the factorization scale was changed 
and the renormalization scale was kept fixed ($\mu_R=M_A, \mu_F=\chi M_A$), while in the one on the right, the opposite variation was performed 
($\mu_F=M_A, \mu_R=\chi M_A$).
As expected from the running of $\as$, the cross section typically decreases when $\mu_R$ increases. This effect is clearly noticeable in the right-side plot, and, more moderate in the left-side plot. The graph on the center shows how the variation of $\mu_F$ leads to and opposite behavior, ie: the cross section increases with the growing of $\mu_F$. 
This can be explained by the following fact: at the LHC the cross section is mainly sensitive to partons with momentum fraction $x \sim 10^{-2}$. In this $x$ range, the scaling violation of the parton densities is moderately positive and therefore one observes an artificially reduced factorization scale dependence. In the left-side plot one sees a partial compensation of the two effects, although the $\mu_R$ variation clearly dominates.
Another interesting feature of this plot is that it shows how sizeable are the higher order corrections. The change when going from LO to NLO is quite large, while the inclusion of the NNLO corrections has a moderate impact. We can consider this fact as a hint for the convergence of the perturbative series.
   \begin{figure}
      \begin{center}
        \epsfig{figure=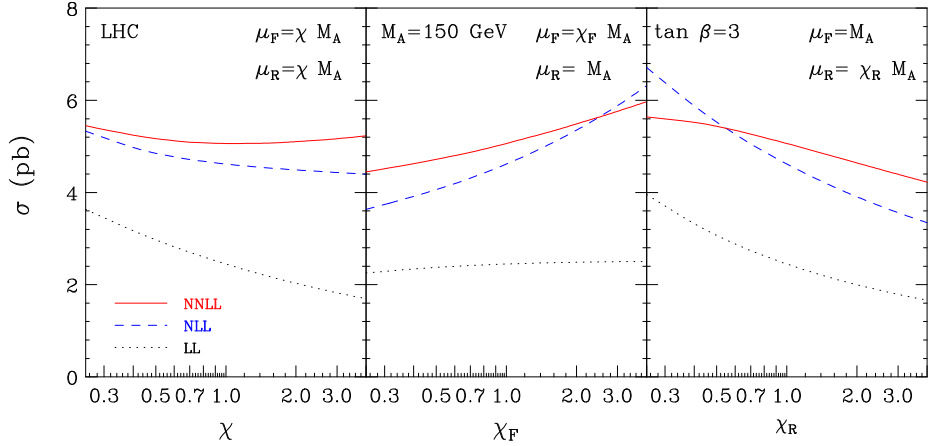,angle=0,height=6cm}
	\end{center}
        \caption{\label{fig:lhc150res}{\em Scale dependence of the Higgs 
production  cross section at the LHC for $M_A=150$ GeV, \tb $=3$, at LL 
(dots), NLL (dashes) and NNLL (solid). }}
     \end{figure}
Fig.~\ref{fig:lhc150res} shows the same as Fig.~\ref{fig:lhc150}, for the resummed cross section.
One sees that the plot on the right keeps the typical dependence on $\mu_R$, due to the  running of the coupling constant. Nevertheless, the behavior for fixed $\mu_R$ has changed, especially if we compare the higher orders (NLO vs NLL and NNLO vs NNLL). The rather flat result of Fig.~\ref{fig:lhc150} is now replaced by a considerably higher variation, due to the inclusion of both soft and collinear higher order terms. Therefore, in the left-side plot, the scale variations are fairly compensated, and particularly the NNLL result exhibits a tenuous dependence on the combined scales. 
Fig.~\ref{fig:lhc150} shows that the scale dependence is very slightly reduced when going from LO to NLO (as was already mentioned), and considerably reduced when going from NLO to NNLO. The implementation of resummation effects leads to a further reduction of the scale dependence.  

\begin{figure}[h]
      \begin{center}
        \epsfig{figure=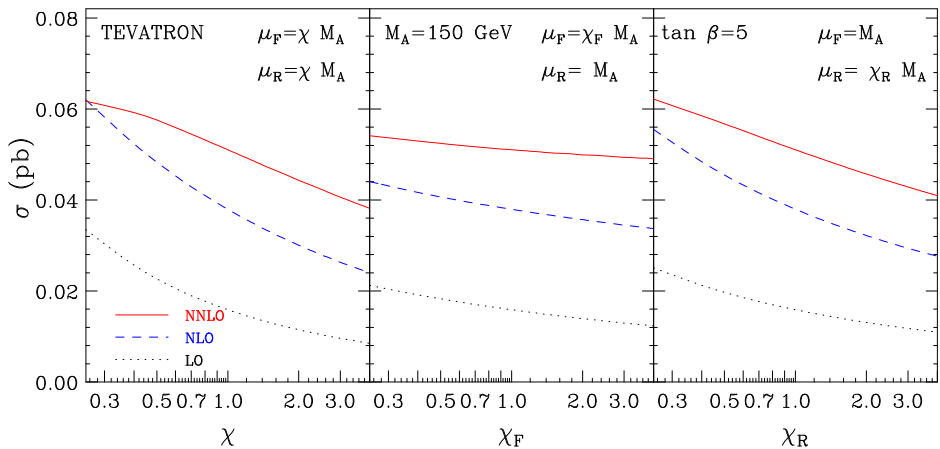,angle=0,height=6cm}
	\end{center}
        \caption{\label{fig:teva150}{\em Scale dependence of the Higgs 
production  cross section at the Tevatron for $M_A$=150 GeV, \tb $= 5$, at 
LO (dots), NLO (dashes) and NNLO (solid). }}
     \end{figure}

\begin{figure}[h]
      \begin{center}
        \epsfig{figure=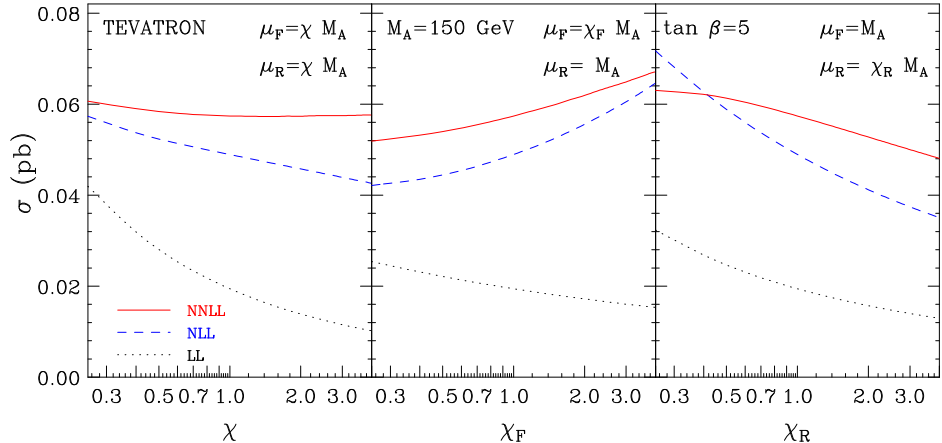,angle=0,height=6cm}
	\end{center}
        \caption{\label{fig:teva150res}{\em Scale dependence of the Higgs 
production  cross section at the Tevatron for $M_A=150$ GeV, \tb $= 5$, at LL 
(dots), NLL (dashes) and NNLL (solid). }}
     \end{figure}

In Fig.~ \ref{fig:teva150}  we present the fixed-order scale dependence 
 for $M_A=150$ GeV with \tb $=5$ now at the Tevatron.
In this Figure, many of the overall features that appeared in the LHC plots are present. The right plot still shows the same dependence, again because of the running of $\as$. In the middle plot, in direct contrast with the result from Fig.~\ref{fig:lhc150},  the cross sections \emph{increases} with $\mu_F$. At the Tevatron, the partons with roughly $x \sim 0.1$ contribute the most and in this region the scaling violation is slightly negative. Then, in the first plot both scales lead to an overall decrease in the cross section. 
Fig.~ \ref{fig:teva150res} shows now the scale dependence of the resummed cross section with the same parameters of the previous figure.
The results are very similar to those in Fig.~\ref{fig:lhc150res}. Here, it is noticeable the rather flat scale dependence of the NNLL result in the left plot. It is quite remarkable that the LO and LL curves look very much alike. The effects of threshold resummation become important only when going to higher (NLL and NNLL) orders.
As an overall feature of the scale dependence graphs, it is important to stress the fact that the resummed cross section is clearly more stable against scale variations than the fixed-order result.

\begin{figure}[h]
      \begin{center}
        \epsfig{figure=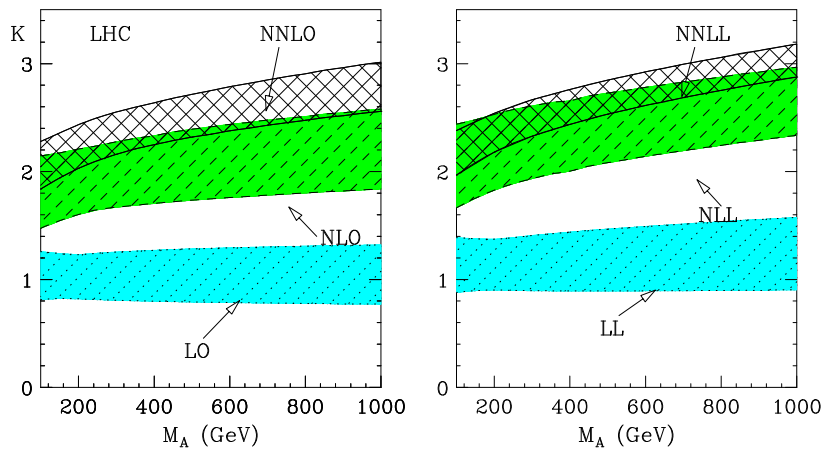,angle=0,height=6cm}
	\end{center}
        \caption{\label{fig:lhck}{\em Fixed-order and resummed K factors 
for Higgs production at LHC.}}
     \end{figure}

The importance of higher-orders effects in commonly presented through the introduction of
K-factors, which are defined as the ratio of the cross section at a given 
order over the LO result.
As it was mentioned before, within the large-$M_t$ approximation, the higher order cross sections are proportional to $\sigma_0$, which is the only term that depends upon \tb. Therefore, in the infinite top mass limit, the K-factors are fully independent on that parameter. In Fig.~\ref{fig:lhck} we 
present the K factor at LHC including its scale dependence. The bands were obtained by 
independently varying the scales in the region 
$0.5  M_A \le \mu_F,\mu_R \le 2 M_A$, with the constraint $0.5 < \mu_F / \mu_R <2$. The LO result that renormalizes the 
K factor was computed with $\mu_F=\mu_R=M_A$.
 The NLO K factor is around $2$, accounting for an increase in the cross sections of approximately the same amount as the LO result itself. The NNLO K factor shows a rather more moderate impact. The inclusion of soft-gluon effects at NLL and NNLL accuracy slightly increases the cross section on top of the fixed-order contributions and show a larger overlap between the corresponding bands, indicating  a better convergence for the resummed series.
It becomes clear the reduction of the scale dependence for the higher orders, as the bands become thinner as the order grows. We also notice an increase of the K factors with $M_A$, consistent with the fact that the soft-gluon contributions become more dominant as the process gets closer to the hadronic threshold. 
Once the resummation is performed, the uncertainty due to scale variation can be estimated to the order of 10 percent. 

\begin{figure}[h]
      \begin{center}
        \epsfig{figure=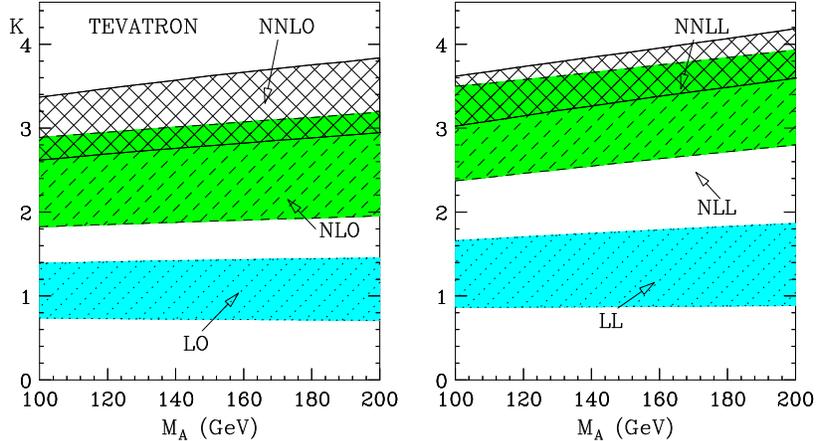,angle=0,height=6cm}
	\end{center}
        \caption{\label{fig:tevak}{\em Fixed-order and resummed K factors 
for Higgs production at Tevatron.}}
     \end{figure}

Finally, Fig.~\ref{fig:tevak} shows the K-factor Higgs mass dependence at the 
Tevatron.
Here we can see the same overall features as presented in Fig.~\ref{fig:lhck}. As a major difference, we note that the K factors are considerably bigger for the Tevatron. This is due to the fact that the Tevatron center of mass energy is closer to the hadronic threshold, which is the kinematical region where soft-gluon effects become relevant.

\section{Conclusions}
\label{sec:conclu}
We have presented the resummed cross section for pseudoscalar 
Higgs production by gluon fusion at NNLL accuracy in hadronic colliders, presenting the most relevant results for the LHC and Tevatron.

Comparing with the fixed-order calculation, the numerical impact of the resummation effects was found to be rather moderate. This slight variation when going from NNLO to NNLL is providing a hint for the (hopefully) faster convergence of the perturbative series. Probably the most striking feature of the resummed result is the considerably reduction of the scale dependence. It allows to make theoretical predictions with a precision of about 10 \%, which are accurate enough for discovery of a Higgs boson at the LHC and Tevatron.
Moreover, our results presents a probe to supersymmetry, as can also be useful to directly test the MSSM and/or another supersymmetric extension of the Standard Model in the low \tb regime.

The Fortran code THIGRES, which computes total cross sections for both scalar
and pseudoscalar Higgs boson up to NNLO (or NNLL), is provided upon request 
from the authors. .

\noindent {\bf  Acknowledgements.} 
This work has been partially supported by ANPCYT, UBACyT and CONICET.


\begin{thebibliography}{99}


\bibitem{Spira:1993bb}
  M.~Spira, A.~Djouadi, D.~Graudenz and P.~M.~Zerwas,
  Phys.\ Lett.\  B {\bf 318}, 347 (1993).

\bibitem{Djouadi:1992pu}
  A.~Djouadi, J.~Kalinowski and P.~M.~Zerwas,
  Z.\ Phys.\  C {\bf 57}, 569 (1993).

\bibitem{Schael:2006cr}
  S.~Schael {\it et al.}  [ALEPH Collaboration],
  Eur.\ Phys.\ J.\  C {\bf 47}, 547 (2006)
  [arXiv:hep-ex/0602042].


\bibitem{Harlander:2006pc}
  R.~Harlander,
  Pramana {\bf 67}, 875 (2006)
  [arXiv:hep-ph/0606095].



\bibitem{Spira:1995rr}
  M.~Spira, A.~Djouadi, D.~Graudenz and P.~M.~Zerwas,
  Nucl.\ Phys.\  B {\bf 453}, 17 (1995)
  [arXiv:hep-ph/9504378].

\bibitem{heavymt}
  J.~R.~Ellis, M.~K.~Gaillard and D.~V.~Nanopoulos,
  Nucl.\ Phys.\  B {\bf 106}, 292 (1976); 
  M.~A.~Shifman, A.~I.~Vainshtein, M.~B.~Voloshin and V.~I.~Zakharov,
  Sov.\ J.\ Nucl.\ Phys.\  {\bf 30}, 711 (1979)
  [Yad.\ Fiz.\  {\bf 30}, 1368 (1979)].


\bibitem{tm}
D.~de Florian, M.~Grazzini and Z.~Kunszt,
Phys.\ Rev.\ Lett.\  {\bf 82} (1999) 5209;
V.~Ravindran, J.~Smith and W.~L.~Van Neerven,
Nucl.\ Phys.\ B {\bf 634} (2002) 247;
C.~J.~Glosser and C.~R.~Schmidt,
JHEP {\bf 0212} (2002) 016.

\bibitem{tmres}
  G.~Bozzi, S.~Catani, D.~de Florian and M.~Grazzini,
  arXiv:0705.3887 [hep-ph];   Nucl.\ Phys.\ B {\bf 737} (2006) 73 
  and references therein.



\bibitem{Harlander:2001is}
  R.~V.~Harlander and W.~B.~Kilgore,
  Phys.\ Rev.\  D {\bf 64}, 013015 (2001)
  [arXiv:hep-ph/0102241].

\bibitem{Catani:2001ic}
  S.~Catani, D.~de Florian and M.~Grazzini,
  JHEP {\bf 0105}, 025 (2001)
  [arXiv:hep-ph/0102227].

\bibitem{Ravindran:2003um}
  V.~Ravindran, J.~Smith and W.~L.~van Neerven,
  Nucl.\ Phys.\  B {\bf 665}, 325 (2003)
  [arXiv:hep-ph/0302135].

\bibitem{Anastasiou:2002yz}
  C.~Anastasiou and K.~Melnikov,
  Nucl.\ Phys.\  B {\bf 646}, 220 (2002)
  [arXiv:hep-ph/0207004].

\bibitem{Harlander:2002wh}
  R.~V.~Harlander and W.~B.~Kilgore,
  Phys.\ Rev.\ Lett.\  {\bf 88}, 201801 (2002)
  [arXiv:hep-ph/0201206].


\bibitem{Anastasiou:2005qj}
  C.~Anastasiou, K.~Melnikov and F.~Petriello,
  Nucl.\ Phys.\  B {\bf 724}, 197 (2005)
  [arXiv:hep-ph/0501130].

\bibitem{Anastasiou:2002wq}
  C.~Anastasiou and K.~Melnikov,
  Phys.\ Rev.\  D {\bf 67}, 037501 (2003)
  [arXiv:hep-ph/0208115].


\bibitem{Harlander:2002vv}
  R.~V.~Harlander and W.~B.~Kilgore,
  JHEP {\bf 0210}, 017 (2002)
  [arXiv:hep-ph/0208096].



\bibitem{Catani:2003zt}
  S.~Catani, D.~de Florian, M.~Grazzini and P.~Nason,
  JHEP {\bf 0307}, 028 (2003)
  [arXiv:hep-ph/0306211].

\bibitem{Dawson:1990zj}
  S.~Dawson,
  Nucl.\ Phys.\  B {\bf 359}, 283 (1991).



\bibitem{Kramer:1998iq}
M.~Kr\"amer, E.~Laenen and M.~Spira,
Nucl.\ Phys.\ B {\bf 511} (1998) 523.

\bibitem{Chetyrkin}
K.~G.~Chetyrkin, B.~A.~Kniehl and M.~Steinhauser,
Phys.\ Rev.\ Lett.\ {\bf 79} (1997) 353, Nucl.\ Phys.\ B {\bf 510} (1998) 61.


\bibitem{Field:2002pb}
  B.~Field, J.~Smith, M.~E.~Tejeda-Yeomans and W.~L.~van Neerven,
  Phys.\ Lett.\  B {\bf 551}, 137 (2003)
  [arXiv:hep-ph/0210369].

\bibitem{Sterman:1986aj}
G.~Sterman,
Nucl.\ Phys.\ B {\bf 281} (1987) 310.


\bibitem{CatTrenres}
  S.~Catani and L.~Trentadue,
  Nucl.\ Phys.\  B {\bf 327}, 323 (1989), Nucl.\ Phys.\ B {\bf 353} (1991) 183.




\bibitem{Moch:2005ba}
  S.~Moch, J.~A.~M.~Vermaseren and A.~Vogt,
  Nucl.\ Phys.\  B {\bf 726}, 317 (2005)
  [arXiv:hep-ph/0506288].

\bibitem{Moch:2005ky}
  S.~Moch and A.~Vogt,
  Phys.\ Lett.\  B {\bf 631}, 48 (2005)
  [arXiv:hep-ph/0508265].


\bibitem{Catani:1996yz}
  S.~Catani, M.~L.~Mangano, P.~Nason and L.~Trentadue,
  Nucl.\ Phys.\  B {\bf 478}, 273 (1996)
  [arXiv:hep-ph/9604351].



\bibitem{Blumlein:2005im}
  J.~Blumlein and V.~Ravindran,
  Nucl.\ Phys.\  B {\bf 716}, 128 (2005)
  [arXiv:hep-ph/0501178].

\bibitem{Mellingredients}
  J.~Blumlein and S.~Kurth,
  Phys.\ Rev.\  D {\bf 60}, 014018 (1999)
  [arXiv:hep-ph/9810241];
  J.~Blumlein and S.~O.~Moch,
  Phys.\ Lett.\  B {\bf 614}, 53 (2005)
  [arXiv:hep-ph/0503188].


\bibitem{Blumlein:2000hw}
  J.~Blumlein,
  Comput.\ Phys.\ Commun.\  {\bf 133}, 76 (2000)
  [arXiv:hep-ph/0003100].


\bibitem{deFlorian:2005yj}
  D.~de Florian and W.~Vogelsang,
  Phys.\ Rev.\  D {\bf 71}, 114004 (2005)
  [arXiv:hep-ph/0501258].


\bibitem{Spira:1995mt}
  M.~Spira,
  arXiv:hep-ph/9510347.


\bibitem{Martin:2001es}
  A.~D.~Martin, R.~G.~Roberts, W.~J.~Stirling and R.~S.~Thorne,
  Eur.\ Phys.\ J.\  C {\bf 23}, 73 (2002)
  [arXiv:hep-ph/0110215].

\bibitem{Martin:2002dr}
  A.~D.~Martin, R.~G.~Roberts, W.~J.~Stirling and R.~S.~Thorne,
  Phys.\ Lett.\  B {\bf 531}, 216 (2002)
  [arXiv:hep-ph/0201127].



\end{thebibliography}
\end{document}